\documentclass[12pt,preprint]{aastex}

\usepackage{xcolor}

\shorttitle{Double-diffusive erosion of the core of Jupiter}
\shortauthors{Moll et al.}

\begin{document}


\title{Double-diffusive erosion of the core of Jupiter}



\author{R. Moll$^{1,2}$, P. Garaud$^2$, C. Mankovich$^3$ and J. J. Fortney$^3$}
\affil{
$^{1}$Department of Oceanography, Naval Postgraduate School, Monterey, CA 93943  \\
$^{2}$Applied Mathematics and Statistics Department, University of California, Santa Cruz, CA 95064,\\
$^{3}$Astronomy \& Astrophysics Department, University of California, Santa Cruz, CA 95064
}



\begin{abstract}
We present Direct Numerical Simulations of the transport of heat and heavy elements across a double-diffusive interface or a double-diffusive staircase, in conditions that are close to those one may expect to find near the boundary between the heavy-element rich core and the hydrogen-helium envelope of giant planets such as Jupiter. We find that the non-dimensional ratio of the buoyancy flux associated with heavy element transport to the buoyancy flux associated with heat transport lies roughly between 0.5 and 1, which is much larger than previous estimates derived by analogy with geophysical double-diffusive convection. Using these results in combination with a core-erosion model proposed by Guillot et al. (2004), we find that the entire core of Jupiter would be eroded within less than 1Myr assuming that the core-envelope boundary is composed of a single interface. We also propose an alternative model that is more appropriate in the presence of a well-established double-diffusive staircase, and find that in this limit a large fraction of the core could be preserved. These findings are interesting in the context of Juno's recent results, but call for further modeling efforts to better understand the process of core erosion from first principles. 
\end{abstract}


\keywords{hydrodynamics --- instabilities --- turbulence --- planets and satellites: gaseous planets --- planets and satellites: interiors}

\section{Introduction}\label{sec:intro}

Recent measurements of the gravitational moments from the Juno mission have been interpreted in the light of structure and evolution models to be less compatible with the existence of a segregated core, and more compatible with a smoother distribution of heavy elements from the center up to 0.3 to 0.5 Jupiter radii \citep{Wahl2017}. If confirmed, this conclusion has one of two fundamental implications for the formation and evolution of Jupiter to the present age. Either, the planet was formed with a more diffuse distribution of heavy elements than previously thought in the context of the standard core-accretion scenario for planet formation \citep{HelledStevenson2017}, {\it or} the planet was assembled via core accretion, but the core was later deeply eroded by the convection above. 

The possibility of core erosion, first raised by \citet{Stevenson1982,Stevenson1985}, has regularly been revisited, notably by \citet{Guillot2004}, and most recently by \citet{Soubiranal2017}. Using new ab-initio calculations of the diffusion coefficients (specifically the viscosity $\nu$ and the molecular diffusivity $\kappa_C$) for a mixture of H, He and traces of SiO$_2$ in conjunction with simple scaling arguments from Rayleigh-B\'enard convection, \citet{Soubiranal2017} argue that the diffusion timescale of heavier species through a laminar boundary layer at the interface between Jupiter's core and convective envelope would be of the order of 10$^6$ years, and conclude that this would erode the core entirely at the present age. Their argument, however, implicitly assumes that the core-envelope interface is a solid surface, and ignores the fact that convection is likely partly suppressed by the strong chemical stratification of the core-envelope interface \citep{Stevenson1982}. 

\begin{figure}[h!]
\includegraphics[width=0.9\textwidth]{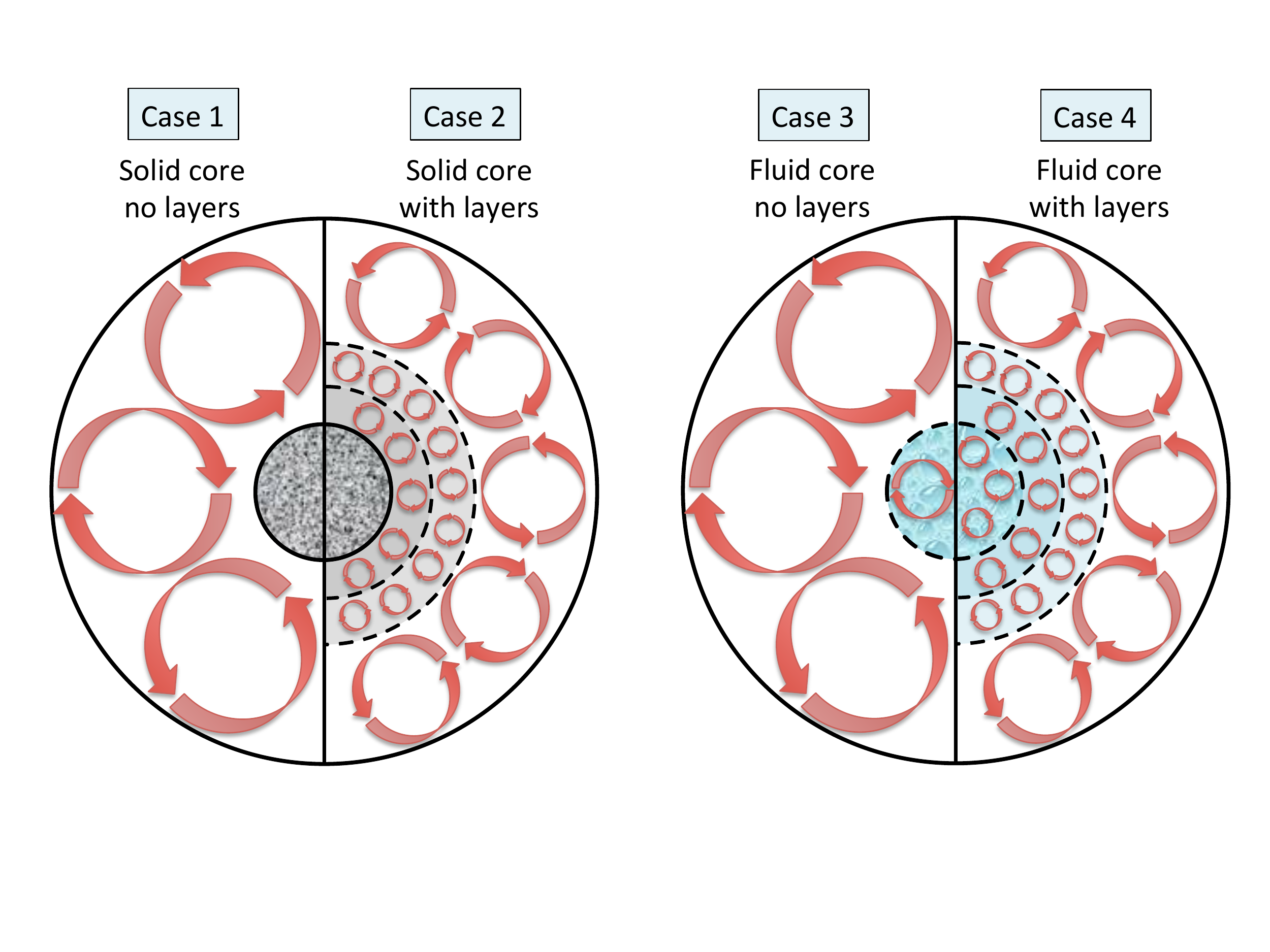}
\caption{Pictorial representation of the nature of the core-envelope interface of a giant planet. Four scenarios are presented. In cases 1 and 2, the core is solid, while in cases 3 and 4, the core is fluid. In cases 2 and 4, a double-diffusive staircase forms outside the core. In all four diagrams, a solid circle (aside from that at the surface) represents a purely diffusive interface between a fluid and a solid, while a dashed circle represents an interface between two fluids, which may or may not be diffusive. Circular arrows depict convective motions. }
\label{fig:cases}
\end{figure}

In fact, dissolution dynamics at the core-envelope interface can take a number of different forms, depending on the nature of the core (solid vs. fluid), and on the compositional stratification just outside the core. Four cases are illustrated in Figure \ref{fig:cases}. In cases 1 and 2, the core is assumed to be in solid form, and is dissolving into the envelope through a narrow laminar boundary layer \citep[case 1 for instance depicts the idea proposed by][]{Soubiranal2017}. In these two cases the eroded mass flux can be estimated provided one knows the convective velocities in the region just above the core. In cases 3 and 4, by contrast, the core is assumed to be in fluid form and likely undergoes some convective mixing. In these cases, the transport across the lowermost interface can be turbulent or diffusive, depending on the nature of that interface (see Section \ref{sec:results} for more on this topic). Interestingly, for Jupiter at least, the exact nature of the core remains uncertain, and the latter could be either in fluid or solid form depending on its (unknown) composition. For instance \citet{French2009} find that water would be a dense plasma under Jovian core conditions, while \citet{WilsonMilitzer2012} find that the rocky material MgO is solid under the same conditions.  Furthermore, \citet{Gonzalesal2014} find that SiO$_2$ is currently solid at Jovian core conditions but would have been liquid when the planet was younger and possessed a hotter interior. As a result, cases 1, 2, 3 and 4 are all possible under different formation scenarios and/or at different stages of the evolution of the planet. 
 
In cases 1 and 3, it is assumed that the eroded material is very rapidly mixed into the envelope, and does not substantially affect convection. As a result, the entire envelope is fully convective and adiabatic. This is what is usually assumed in planetary structure and evolution models. However, as discussed by \citet{Stevenson1985}, depending on the formation history of the planet a substantial compositional gradient could exist above the core and partially inhibit convection. In that case, and by analogy with geophysical thermo-compositional convection,  a series of double-diffusive layers could\footnote{Note that another possibility exists, in which the double-diffusive region with strong compositional stratification adjacent to the core is the seat of weakly nonlinear oscillatory double-diffusive convection (ODDC) instead of layered convection. In that case, as shown by \citet{Moll2016}, the turbulent transport of both heat and composition would be close to diffusive across the entire region. However, whether this quiescent state could exist in such close proximity to a strong convective zone is questionable.} develop \citep[e.g.][]{TURNER1965,HuppertLinden79}. These are shown in cases 2 and 4. The interfaces between the layers can be turbulent or diffusive, depending on their nature. In this kind of layered convection, the vertical extent of convective eddies is limited to the height of the layers, which implies that both turbulent fluxes of heat and composition are much reduced compared with the fully-convective case. This has in fact also been proposed as a possible mechanism to slow down the cooling of giant planets \citep{Stevenson1985,LeconteChabrier2012}.

In all but case 1, we therefore see that understanding transport through one (case 3) or several double-diffusive interfaces (cases 2 and 4) is crucial if one wishes to study the fate of Jupiter's core. A first quantitative estimate of the core erosion rate through a single double-diffusive interface (e.g. case 3) was put forward by \citet{Guillot2004}. They did so by using a well-known result on the transport of heat and chemical species across such interfaces in laboratory experiments, first performed by \citet{TURNER1965} (heat/salt interface) and \citet{Shirtcliffe1973} (salt/sugar interface, where salt plays the role of heat), and modeled by \citet{LindenShirtcliffe1978}. These experiments and subsequent modeling efforts \citep[see the review by][]{fernando1989buoyancy} suggest that, when the compositional jump across the interface is sufficiently large (which would be the case in an initially differentiated planet), the ratio of the buoyancy flux associated with chemical transport across the interface ${\cal F}_C$, to the equivalent buoyancy flux associated with heat transport ${\cal F}_T$, is a constant\footnote{The quantity is defined as an inverse for consistency with prior work going back to \citet{radko2003mechanism}, which defines $\gamma$ as the ratio of the buoyancy flux associated with thermal transport to that associated with compositional transport.} $\gamma^{-1}$ that only depends on the ratio of the chemical to thermal diffusivity in the system. Mathematically speaking, \citet{LindenShirtcliffe1978} argue that 
\begin{equation}
\gamma^{-1} = \frac{{\cal F}_C}{{\cal F}_T} = \left( \frac{\kappa_C}{\kappa_T} \right)^{1/2} \equiv \tau^{1/2} 
\end{equation}
where $\kappa_T$ is the local thermal diffusivity. The diffusivity ratio $\tau = \kappa_C/ \kappa_T$ is very small in the core of Jupiter, and has been estimated to be of order $10^{-2}$  which would imply $\gamma^{-1} \sim 0.1$ \citep{Stevenson1982}. \citet{Guillot2004} used this estimate for the flux ratio $\gamma^{-1}$ as a proxy for the ratio of the potential to thermal energy flux carried by large-scale convective eddies, and thus derived the following equation for the core erosion rate: 
\begin{equation}
\frac{dm_{\rm core}}{dt} = - \frac{ \gamma^{-1} }{\varpi } \frac{L_1 R}{GM} 
\label{eq:guillot}
\end{equation}
(see their equation 14) where $G$ is the gravitational constant, $R$ and $M$ are the radius and mass of Jupiter at time $t$, $L_1$ is the luminosity 1 pressure scaleheight above the core at the same time, and $\varpi$ is a constant they argue is about 0.3. With $\gamma^{-1} \sim 0.1$, they conclude that convection could erode between 15 and 20 Earth masses ($M_E$ hereafter) from the core of Jupiter (but only a few $M_E$ from the core of Saturn) in 5Gyr.

However, as reviewed by Garaud (2018), double-diffusive dynamics are fundamentally different in geophysical and astrophysical conditions, shedding doubt on the relevance of the \citet{LindenShirtcliffe1978} model to estimate $\gamma^{-1}$ in the interior of Jupiter. Indeed, one of the main parameters governing the dynamics of fluids is the Prandtl number $\Pr = \nu/\kappa_T$. The latter is typically significantly greater than one in geophysical flows, while it is asymptotically small at astrophysical parameters (from $\Pr \sim 10^{-2}$ in the core of giant planets, down to $\Pr \sim 10^{-6}$ in stellar interiors). This has a crucial implication for fluid stability in regions that are stably stratified in composition, but unstably stratified in temperature. Indeed, stability analysis \citep{Walin1964} reveals that the linearly unstable range of parameter space is extremely narrow when $\Pr$ is greater than one, but broadens to include nearly all astrophysically realizable thermocompositional stratifications when $\tau \le \Pr \ll 1$. This implies that almost none of what is known from geophysical double-diffusive convection applies to low Prandtl number astrophysical fluids, starting from the mechanisms by which interfaces (or whole staircases) are formed \citep{rosenblumal2011}, to the dynamical properties of each interface, and ultimately to the global evolution of a system undergoing layered convection \citep{Woodal13}. 

Because of this, new experiments to measure the flux ratio across double-diffusive interfaces at low Prandtl number are required. These must necessarily be numerical in nature, since 
fluids with $\Pr \le 0.1$ are extremely rare on Earth. The first results on this topic were presented by \citet{Woodal13} who ran a series of numerical experiments of double-diffusive convection in the layered regime (with one or multiple interfaces spontaneously forming) with $\Pr$ ranging from 0.01 to 0.3. They focussed on a region of parameter space where the stabilizing compositional stratification (as measured by the inverse density ratio, see Section 2) is relatively weak, as it is the case for instance near the convective core of intermediate-mass stars, and found that $\gamma^{-1}$ is substantially larger than $\tau^{1/2}$ owing to the intrinsically turbulent nature of the interfaces in that regime. 

However, the mean compositional gradient around the core-envelope interface of giant planets is likely much larger than in stars, so the results of \citet{Woodal13} do not directly apply, and additional simulations are required to extend their original study. Furthermore, \citet{Mirouh2012} showed that layered convection cannot arise spontaneously in strongly stratified systems. Layers can nevertheless be present, either as a legacy of initial conditions (e.g. if the planet starts in a chemically differentiated state), or, through other possible pathways. The latter have not really been investigated nor discussed in planetary contexts yet, but known ones in geophysical applications include mechanical mixing by shear or other processes \citep{Veronis1965,Radko2016}, instantaneous bottom heating as in the experiments of \citet{HuppertLinden79}, or through interactions with lateral intrusions, as in the Arctic ocean \citep{BebievaTimmermans2017}. 
 
In what follows, we therefore assume that the giant planet is initially chemically segregated, and contains one or more double-diffusive interfaces separating fully convective layers. We present our model in Section \ref{sec:model}. This model is used in Section \ref{sec:results} to run a series of numerical experiments to measure the flux ratio $\gamma^{-1}$ as a function of various input parameters, such as the Prandtl number $\Pr$, the diffusivity ratio $\tau$, the strength of the compositional stratification, and the overall height of the convective layers. As we shall demonstrate, and not unexpectedly given the discussion above, the behavior of such double-diffusive interfaces is very different from what is known in geophysical systems, and the flux ratio $\gamma^{-1}$ through the core-envelope interface(s) can be substantially larger than $\tau^{1/2}$. In Section \ref{sec:Jupitermodels}, we use our new experimentally derived estimate for $\gamma^{-1}$ in a one-dimensional evolutionary model of Jupiter, and revisit the predictions of \citet{Guillot2004} for the core erosion rate {\color{red} in the context of a Case 3 type of model. We also propose and apply a new model that is more appropriate for transport through a staircase (as in Cases 2 and 4)}. In Section \ref{sec:conclusion}, we discuss the caveats of our findings, and what potential implications they may have to constrain the formation and evolution history of Jupiter in the light of Juno's existing and anticipated gravitational moments measurements. 

\section{Numerical model setup}
\label{sec:model}

In order to study numerically the transport of heavy elements from the core up into the envelope across a single or a series of double-diffusive interfaces, we use a model setup that is similar to the one introduced by \citet{rosenblumal2011} \citep[see also][]{Mirouh2012,Woodal13}. We focus on a small region of the planet that is located in the vicinity of the interface(s), and model it using a local Cartesian domain $(x,y,z)$ with gravity aligned with the vertical axis: ${\bf g} = -g{\bf e}_z$. Rotation is ignored for now, although it is worth noting that the latter could be important in this context \citep{MollGaraud2017}. The height of the domain $L_z$ is assumed to be smaller than any density or temperature scaleheight to allow for the use of the Boussinesq approximation for weakly compressible gases \citep{SpiegelVeronis1960}. This approximation is reasonable in this region of the planet. For simplicity, $g$ and all the diffusivities ($\nu$, $\kappa_T$ and $\kappa_C$) are assumed constant within the domain. We then express the temperature and composition fields as the sum of a linear background stratification plus perturbations,
\begin{eqnarray}
T(x,y,z,t) & = & T_0(z) + \tilde{T}(x,y,z,t), \\
C(x,y,z,t) & = & C_0(z) + \tilde{C}(x,y,z,t),
\end{eqnarray}
where $T_0(z)=T_m+T_{0z}z$ and $C_0(z)=C_m + C_{0z}z$, where $T_m$, $C_m$, $T_{0z}$ and $C_{0z}$ are constant. The compositional field $C$ can be viewed as the concentration of a particular species per unit mass, or the total concentration of heavy elements per unit mass (in which case it can be identified with the usually defined $Z$). It could also be interpreted as the mean molecular weight $\mu$, although we prefer not to do so here for ease of interpretation of the meaning of the fluxes later (see Section \ref{sec:Jupitermodels}). Note that {\color{red} $L_z T_{0z}$, $L_z C_{0z}$,} $\tilde{T}$ and $\tilde{C}$ must be small compared with $T_m$ and $C_m$ to be consistent with the Boussinesq approximation. This allows us to linearize the equation of state: 
\begin{equation}
\frac{\tilde{\rho}}{\rho_m} = -\alpha \tilde{T} + \beta \tilde{C}, \label{eq:eos}
\end{equation}
where $\rho_m = \rho(T_m,C_m)$ is the mean density of the region, and where $\alpha= -\rho_m^{-1} \frac{\partial \rho}{\partial T}$, and $\beta= \rho_m^{-1} \frac{\partial \rho}{\partial C}$ are the corresponding partial derivatives of the equation of state at $T_m$, $C_m$ and $\rho_m$. 
{\color{red} On the other hand $\tilde{T}$ and $\tilde{C}$} can be of the same order as $T_{0z}L_z$ or $C_{0z} L_z$ and are allowed to have a non-zero horizontal mean (this is indeed required in order to model a stationary interface or staircase). 

The nondimensional equations governing the fluid evolution are then \citep{radko2013double}:
\begin{eqnarray}
\frac{1}{\mathrm{Pr}} \left(\frac{\partial{\bf \tilde{u}}}{\partial t} + {\bf \tilde{u}}\cdot\nabla {\bf \tilde{u}}\right) & = & -\nabla \tilde{p} + (\tilde{T} - \tilde{C}){\bf e}_z + \nabla^2 {\bf \tilde{u}} \label{eq:momentum}, \\
\nabla \cdot {\bf \tilde{u}} &=& 0 \label{eq:continuity},  \\
\frac{\partial \tilde{T}}{\partial t} + {\bf \tilde{u}}\cdot\nabla \tilde{T} - \tilde{w}  & = & \nabla^2 \tilde{T} \label{eq:heat}, \\
\frac{\partial \tilde{C}}{\partial t} + {\bf \tilde{u}}\cdot\nabla \tilde{C} - R_0^{-1} \tilde{w} & = & \tau \nabla^2 \tilde{C} \label{eq:composition},
\end{eqnarray}
where $\Pr$, $\tau$ were defined in Section \ref{sec:intro}, and where $R_0^{-1}$ is the inverse density ratio defined here as
\begin{equation}
R_0^{-1} = \frac{\beta C_{0z}}{\alpha (T_{0z} - T^{\rm ad}_z)} ,
\end{equation}
where $T^{\rm ad}_z = - g/ c_p$ is the (constant) adiabatic temperature gradient of the region ($c_p$ being the specific heat at constant pressure).  To nondimensionalize these equations, we have adopted the commonly used unit system for oscillatory double-diffusive convection (ODDC) \citep{radko2013double} in which $d=(\kappa_T\nu/\alpha g|T_{0z}-T^{\rm ad}_z|)^{1/4}$ is the unit length, $d^2/\kappa_T$ is the unit time, $d|T_{0z}-T^{\rm ad}_z|$ is the unit temperature and $(\alpha/\beta)d|T_{0z}-T^{\rm ad}_z|$ is the unit composition. In all that follows we solve (\ref{eq:momentum}--\ref{eq:composition}) in a computational domain of size $(L_x,L_y,L_z)$, assuming that perturbations are triply periodic, namely 
\begin{equation}
\tilde{T}(x,y,z,t) = \tilde{T}(x+L_x,y,z,t) = \tilde{T}(x,y+L_y,z,t) = \tilde{T}(x,y,z+L_z,t),
\end{equation}
and similarly for $\tilde{p}$, $\tilde{C}$ and $\tilde{\bf u}$. All the simulations presented are obtained using the PADDI code described in \citet{Traxler2011a}. 

\citet{Walin1964} showed that this set of equations is linearly unstable to oscillatory double-diffusive convection (ODDC) when $1 < R_0^{-1} < R_c^{-1}$ where
\begin{equation}
R_c^{-1} \equiv \frac{{\rm Pr}+1}{{\rm Pr} + \tau} = \frac{\nu + \kappa_T}{\nu + \kappa_C}.
\end{equation}
Furthermore, as discovered by \citet{rosenblumal2011} and \citet{Mirouh2012} following the work of \citet{radko2003mechanism}, ODDC is unstable to a secondary layering instability at low inverse density ratios, for $1<  R_0^{-1} < R_L^{-1} < R_c^{-1}$. This instability causes the spontaneous formation of a thermocompositional staircase in which well-mixed convective layers are separated by stably stratified albeit turbulent interfaces. The critical density ratio for spontaneous layering was numerically determined to be $R_L^{-1} \simeq 3$ when $\Pr = \tau = 0.01$ \citep{Mirouh2012}. The mean inverse density ratio near the core-envelope interface of giant planets being typically larger than 3 but much smaller than $ R_c^{-1}$ (which is $O(10^2)$ if $\Pr \sim \tau \sim O(10^{-2})$), a staircase is not expected to form spontaneously from an initially linear background stratification. As a result, double-diffusive layering in giant planets can take the form of a single interface resulting from the original formation of the planet, or, a set of interfaces that could have appeared as a result of so-far unspecified mechanical or thermal processes (see Section \ref{sec:intro}). To model layers numerically in the desired regime $ R_L^{-1} \ll R_0^{-1} \ll R_c^{-1}$, we must therefore impose them as a initial conditions. 

In all of the simulations presented below, the total temperature and composition fields $\tilde{T} + T_0(z)$ and $\tilde{C} + C_0(z)$ are initialized with small amplitude random noise on top of a $z$-dependent hyperbolic tangent profile of thickness $h$, which mimics the overall profile of a fully-formed staircase, with layers of neutral buoyancy separated by thin, stably stratified interfaces. The value of $h$ is somewhat irrelevant since the staircase later adjusts to a unique statistically-stationary state. For simplicity, we have therefore chosen $h = 15d$ in all simulations. Note that we have tested other sets of initial conditions where $T$, $S$ and $\rho$ have different functional forms but the same number of interfaces in the same domain height. As long as the interfaces are equally spaced, the final statistically stationary state reached by the staircase is also usually the same (unless mergers take place, see Garaud et al. in prep.). Finally, we have also confirmed numerically that simulations in domains of size $nL_z$ with $n$ equally-spaced interfaces have the same global properties per unit volume (e.g. heat and compositional flux, kinetic energy, etc..) as simulations with one interface in a domain of size $L_z$. For this reason, we restrict the presentation of our results below to simulations that only contain one interface in a domain of size $L_z$, bearing in mind that this is actually equivalent to an infinite periodic staircase with constant step-size $L_z$. 

\section{Results}
\label{sec:results}

\subsection{Qualitative results} 
\label{subsec:qualres}

\begin{figure}[h!]
\includegraphics[width=0.9\textwidth]{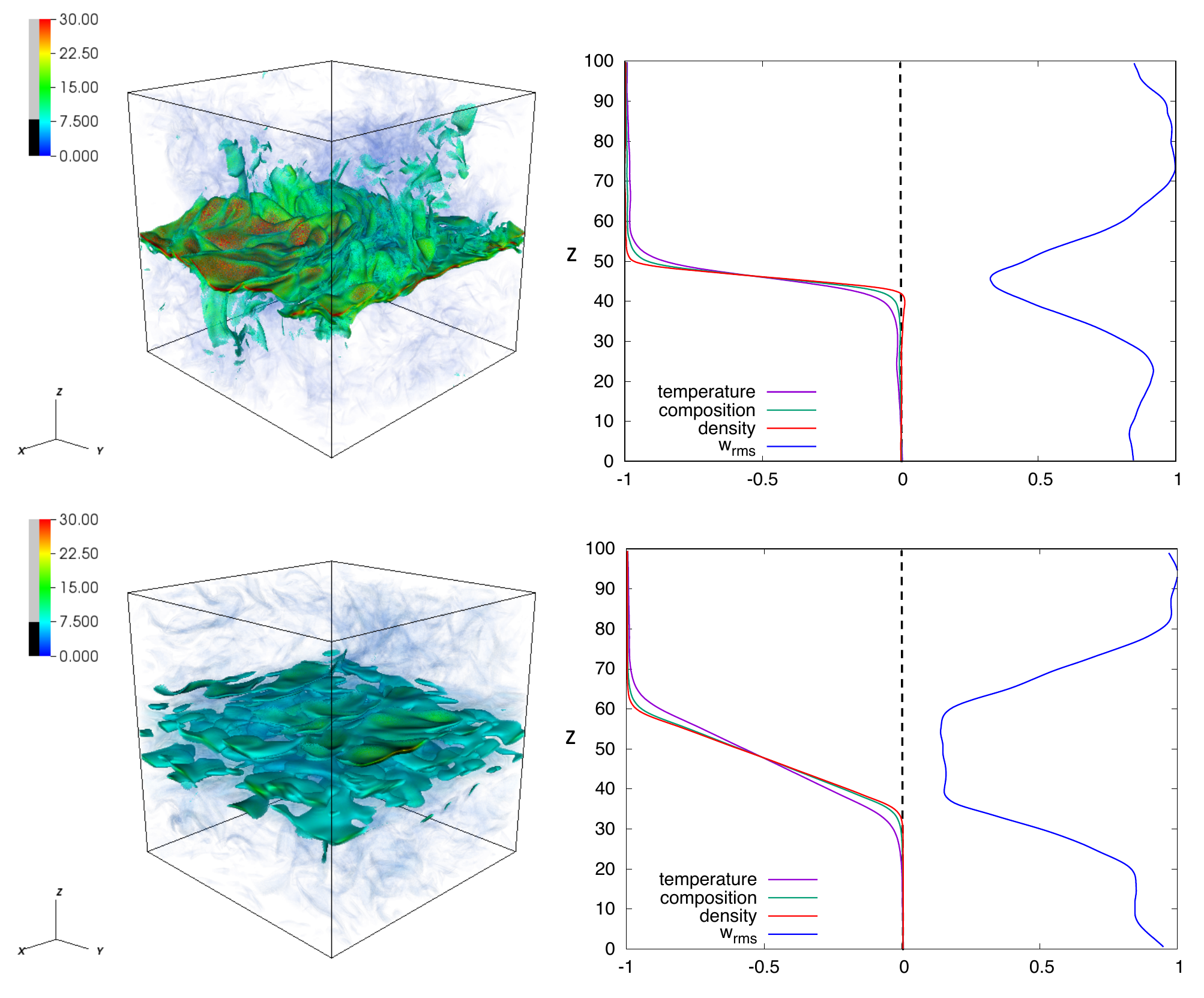}
\caption{Snapshots of simulations at $\Pr = \tau = 0.1$, and $R_0^{-1} = 2.25$ (top row) and $R_0^{-1} = 4.5$ (bottom row), in simulations with $L_x = L_y = L_z= 100d$. Note that $R_L^{-1} \simeq 1.5$ and $R_c^{-1} = 5.5$ at these parameters. Left: Volume rendering of {\color{red} $|\nabla \tilde{C}|$}, the magnitude of the gradient of the compositional {\color{red} perturbations}. Right: Normalized horizontally averaged {\color{red} total} temperature, {\color{red} total} composition, {\color{red} total} density, and rms vertical velocity profiles taken at the same times as the snapshots. }
\label{fig:snaps}
\end{figure}

Our numerical investigation reveals that there are two qualitatively distinct regimes for relatively ``low" and ``high" inverse density ratios (by low and high, we imply $R_0^{-1}$ closer to $R_L^{-1} $ vs.  $R_0^{-1}$ closer to $R_c^{-1}$).  At low $R_0^{-1}$ we observe that the interfaces are very thin but also very mobile. They oscillate up and down as they are pummeled by strong upflows from the lower layer, and downflows from the upper layer. They are also scoured and occasionally pierced by stronger eddies, in what appears to be the main source of transport from one layer to the next. This is similar to what \citet{Woodal13} found in the case of spontaneously emerging layers at even lower $R_0^{-1}$. Typical features of these lower {\color{red} inverse} density ratio interfaces are presented in Figure \ref{fig:snaps} (top row). 

At higher $R_0^{-1}$, the interfaces thicken significantly. They are also much less mobile, with typical variations of the position of the interface being much smaller than its thickness. Transport across the interface is closer to being diffusive (see more on this below) especially at lower values of $\Pr$ and $\tau$, but the interface is not quiescent, as measured for instance using the r.m.s. vertical velocity $w_{\rm rms}$. Instead, it appears to be the seat of weakly nonlinear ODDC \citep{Moll2016}. Typical features of these higher density ratio interfaces are presented in Figure \ref{fig:snaps} (bottom row). 

Our qualitative numerical findings on the nature of double-diffusive interfaces at low Prandtl number are therefore somewhat different from what is assumed in the staircase models of 
\citet{LeconteChabrier2012} and \citet{Spruit2013}. An in-depth analysis of the transport of heat through double-diffusive staircases in the light of these analytical models will be presented elsewhere (Garaud et al., in preparation).

Finally, note that we have not been able to find any statistically stationary layered solutions for $R_0^{-1} > R_c^{-1}$. In fact, for some simulations in smaller computational domains, layered solutions {\color{red} already} disappear for $R_0^{-1}$ somewhat smaller than $R_c^{-1}$. The interfaces of simulations initialized at these high inverse density ratios slowly thicken until they fill the entire domain at which point the layers disappear. This is very different from the oceanographic case, where layered solutions are found to exist for $R_0^{-1}$ significantly greater than $R_c^{-1}$.

\subsection{Flux ratio}\label{subsec:smallflux}

\begin{figure}[h!]
\includegraphics[width=7cm]{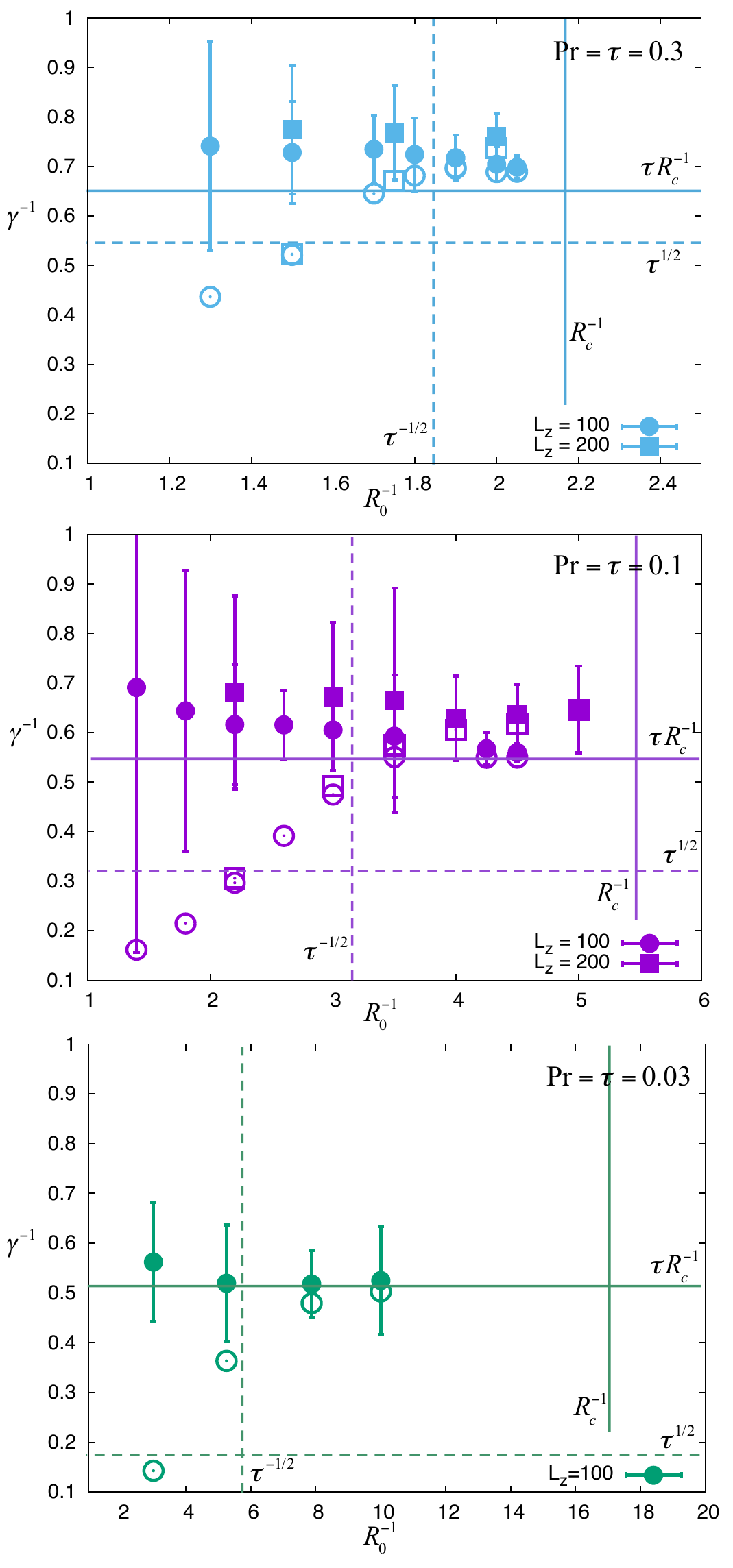}
\caption{Values of $\gamma^{-1}$ measured in simulations with $\Pr = \tau = 0.3$ (top), $\Pr = \tau = 0.1$ (middle) and $\Pr = \tau = 0.03$ (bottom). In each panel, the round symbols are used for runs that have a domain height $L_z = 100d$, while the squares are used for $L_z = 200d$. In all cases, $L_x = L_y = L_z$. Filled symbols show the value of the total flux ratio $\gamma^{-1}$ given by equation (\ref{eq:gammadef}), while open symbols show the diffusive flux ratio $\gamma^{-1}_{\rm diff}$ of equation (\ref{eq:gammadiff}). Also shown is the theoretical prediction $\gamma^{-1} = \tau^{1/2}$ of \citet{LindenShirtcliffe1978} (horizontal dashed line), as well as the maximum density ratio $\tau^{-1/2}$ where stable staircases are supposed to exist in their model (vertical dashed line). }
\label{fig:gamma}
\end{figure}

Of particular interest in the context of the core erosion process for giant planets is the flux ratio $\gamma^{-1}$ defined in Section \ref{sec:intro}. We have measured it in all the 1-layer simulations available to date, using the following method. A simulation initiated in a 1-layered state is evolved until a statistically stationary state is reached. Once a sufficiently long time series is available in that state, we measure the volume- and time-averaged $\gamma^{-1}$ from the ratio of the total nondimensional composition flux to the total nondimensional temperature flux:
\begin{equation}
\gamma^{-1} = \frac{ \tau R_0^{-1} + \langle \tilde{w}\tilde{C} \rangle }{1 + \langle \tilde{w}\tilde{T}\rangle} = \tau \frac{  R_0^{-1} + R_0  \langle | \nabla \tilde{C}|^2 \rangle }{1 + \langle | \nabla \tilde{T}|^2\rangle},
\label{eq:gammadef}
\end{equation}
where the angular brackets denote a volume and time average, and all quantities are implicitly nondimensional unless otherwise noted. To arrive at the last expression above, we have used the fact that, in a statistically stationary state, 
\begin{equation}
\langle \tilde{w}\tilde{T}\rangle = \langle | \nabla \tilde{T}|^2\rangle  \mbox{ and   } R_0^{-1}\langle \tilde{w}\tilde{C} \rangle= \tau  \langle | \nabla \tilde{C}|^2 \rangle .
\end{equation}
This substitution is done because the dissipation terms $\langle | \nabla \tilde{T}|^2\rangle$ and $\langle | \nabla \tilde{C}|^2\rangle$ have smaller rms fluctuations than the turbulent fluxes $\langle \tilde{w}\tilde{T}\rangle$ and $\langle \tilde{w}\tilde{C}\rangle$, which oscillate in response to the internal gravity waves associated with ODDC \citep{Woodal13}. The {\color{red} values of $\gamma^{-1}$ thus measured} are shown in Figure \ref{fig:gamma} {\color{red} as a function of $R_0^{-1}$, $\Pr$ and $\tau$}. 

We can immediately draw a number of conclusions, first and foremost that using $\tau^{1/2}$ \citep{LindenShirtcliffe1978} severely underestimates $\gamma^{-1}$ in low Pr and high $R_0^{-1}$ layered convection, especially for the lowest values of $\Pr$ and $\tau$. Secondly, we see that layers can exist for  $R_0^{-1}$ significantly greater than $\tau^{-1/2}$, which would be the {\color{red} predicted} upper limit in the high Prandtl number case \citep{LindenShirtcliffe1978} (also see Section \ref{subsec:why}). We can also deduce from Figure \ref{fig:gamma} 
that $\gamma^{-1}$ depends on the layer height in a periodic staircase. Finally, we clearly see the existence of the two regimes suspected in Section \ref{subsec:qualres}: at lower $R_0^{-1}$, the flux ratio is much larger than its diffusive contribution $\gamma^{-1}_{\rm diff}$, while for large $R_0^{-1}$, the flux ratio is very close to $\gamma^{-1}_{\rm diff}$, suggesting that the transport through the interface is close to being diffusive (even though turbulent motions are present). Note that $\gamma_{\rm diff}^{-1}$ is calculated by measuring the temperature gradient $dT_I/dz$ and the compositional gradient $dC_I/dz$ in the middle of the interface:
\begin{equation}
\gamma^{-1}_{\rm diff} = \frac{\tau \frac{dC_I}{dz}}{\frac{dT_I}{dz}} \equiv \tau R_I^{-1} 
\label{eq:gammadiff}
\end{equation} 
where $R_I^{-1}$ thus defined is the interfacial inverse density ratio. 

\subsection{Why are low Pr staircases different from high Pr ones?}\label{subsec:why}

In hindsight, this fundamental difference between high and low $\Pr$ staircases could easily have been foreseen, and arises from the simple and easily verified algebraic result that
\begin{equation}
\tau^{-1/2} >  \frac{\Pr+ 1}{\Pr + \tau}  \mbox{   if    } \Pr   > 1 \mbox{   but  }  \tau^{-1/2} < \frac{\Pr+ 1}{\Pr + \tau}  \mbox{   if    } \Pr   < 1 
\label{eq:ineqs}
\end{equation}
for any $\tau  < 1$ (which is true by definition). 

Indeed, at high Prandtl number \citet{LindenShirtcliffe1978} argued that $\gamma^{-1} = \tau^{1/2}$ based on the ratio of the convective fluxes through the layers. In a stationary state, that ratio must be the same as the ratio of the fluxes through the interface. Assuming that the interface is purely diffusive (as they do), implies that 
\begin{equation}
\gamma^{-1} = \tau^{1/2} = \gamma^{-1}_{\rm diff} = \tau R_I^{-1}, 
\end{equation}
which in turn implies that the inverse density ratio of the interface is $R_I^{-1} = \tau^{-1/2}$. Since, by the inequality (\ref{eq:ineqs}), $\tau^{-1/2} > R_c^{-1}$ when $\Pr > 1$, the interface is stable to ODDC, and acts as a very strong stabilizing barrier to any fluid motion \citep[e.g.][]{Carpenter2012}. Finally, since the interfacial density ratio must be larger than the background density ratio $R_0^{-1}$, the fact that $R_I^{-1}$ must be equal to $\tau^{-1/2}$ also implies that no layered solutions can be found for $R_0^{-1} > \tau^{-1/2}$, as mentioned above.

\begin{figure}[h]
\centerline{\includegraphics[width=10cm]{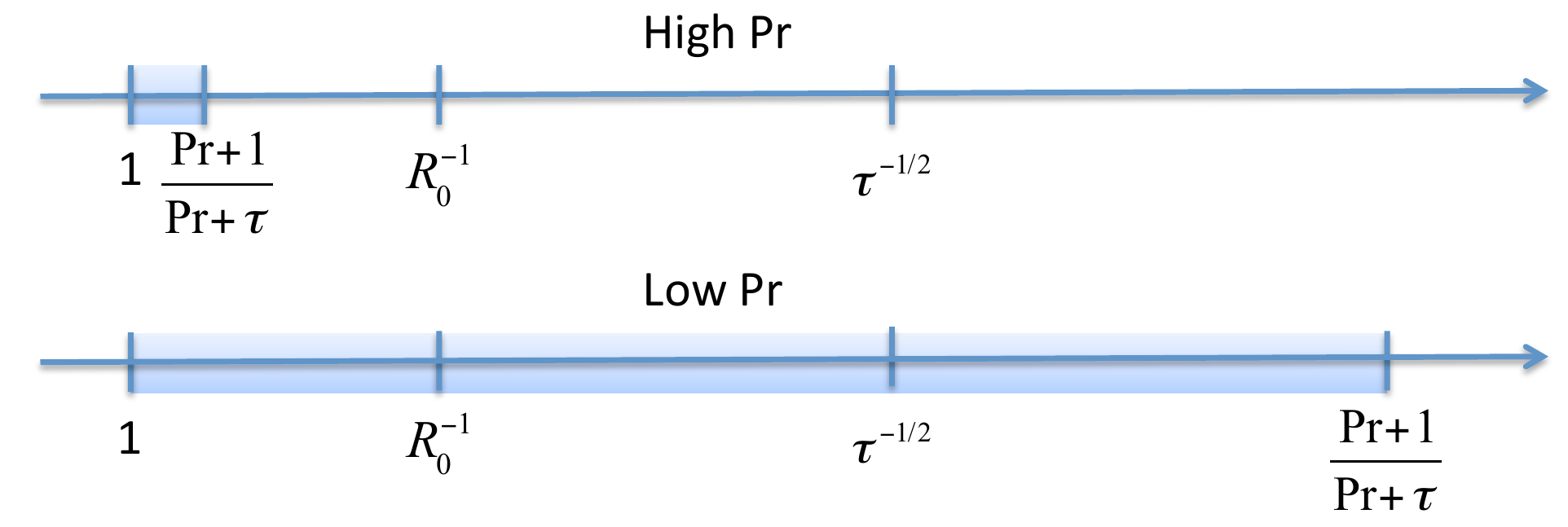}}
\caption{Hierarchy of the important inverse density ratios at high and low Prandtl number. The shaded area is linearly unstable to ODDC. In the Linden and Shirtcliffe model, $R_I^{-1} = \tau^{-1/2}$. This is larger than $R_c^{-1} = (\Pr+1)/(\Pr+\tau)$ and therefore stable when $\Pr > 1$. When $\Pr < 1$ on the other hand $ \tau^{-1/2}<R_c^{-1}$ so the interface is  unstable.}
\label{fig:LS}
\end{figure}

At low Prandtl number on the other hand assuming $\gamma^{-1} = \tau^{1/2}$ leads to an inconsistency in the argument. Indeed, an interface with $R_I^{-1} = \tau^{-1/2}$ cannot be laminar since $\tau^{-1/2} < R_c^{-1}$ when $\Pr < 1$. Instead, it is unstable to ODDC as seen in Figure \ref{fig:snaps}. The hierarchy between $R_0^{-1}$, $R_c^{-1}$ and the assumed value of $R_I^{-1} = \tau^{-1/2}$ in the Linden and Shirtcliffe model, at high and low Pr, is depicted in Figure \ref{fig:LS}.

\begin{figure}[h]
\centerline{\includegraphics[width=10cm]{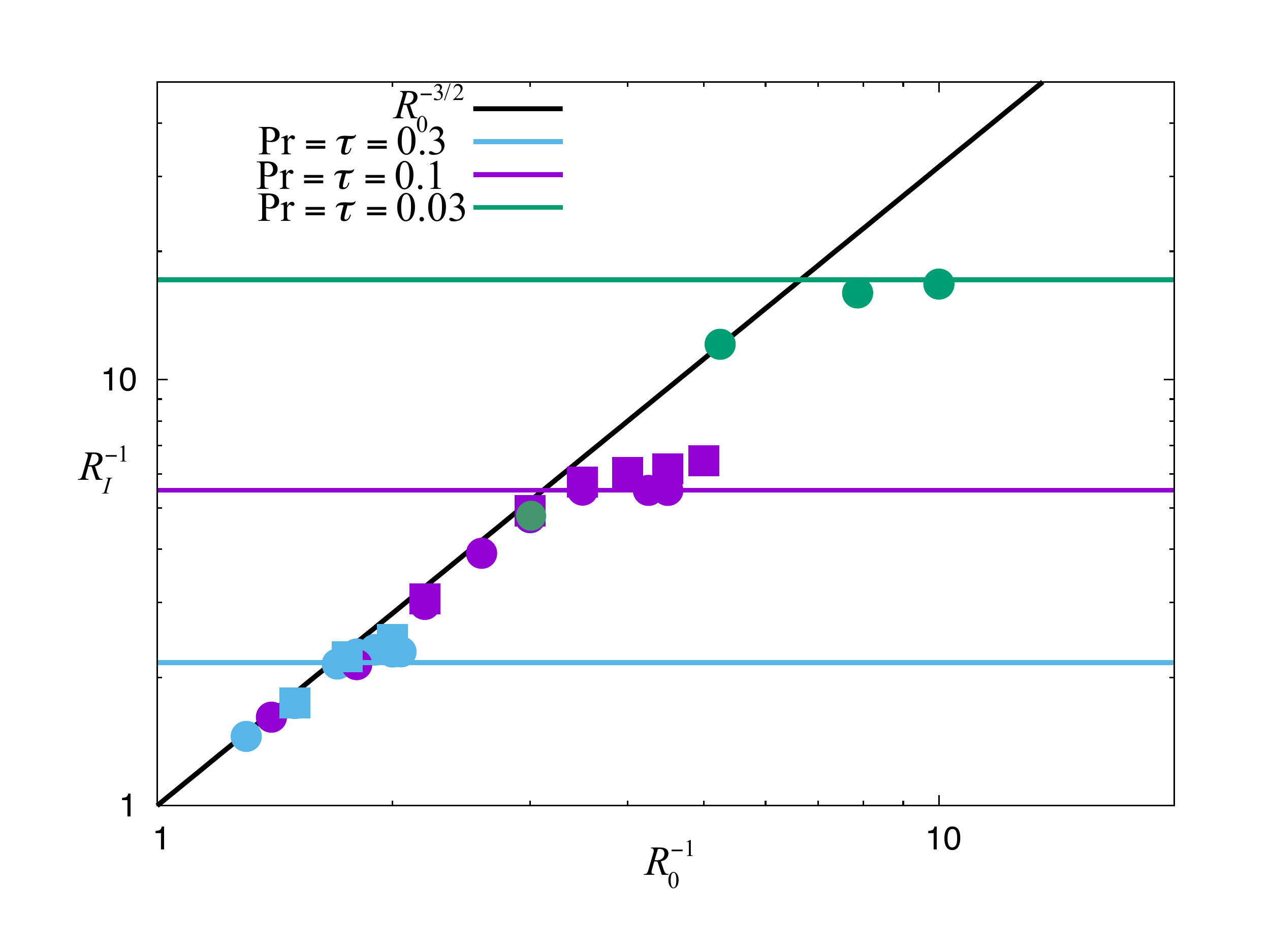}}
\caption{Values of the interfacial inverse density ratio $R_I^{-1}$ measured in our simulations. In each panel, the round symbols are used for runs that have a domain height $L_z = 100d$, while the squares are used for $L_z = 200d$. In all cases, $L_x = L_y = L_z$. The horizontal lines show $R_c^{-1} = (\Pr+1)/(\Pr + \tau)$ for $\Pr = \tau = 0.3$ (blue line), $\Pr = \tau = 0.1$ (purple line) and $\Pr = \tau  = 0.03$ (green line). The slanted black line shows the empirical fit to the low stratification results, $R_I^{-1} \simeq R_0^{-3/2}$.}
\label{fig:Ri}
\end{figure}

To see what the interface adjusts itself to in the low Prandtl number limit, we show the value of $R_I^{-1}$ measured in all of our single-interface simulations in Figure \ref{fig:Ri}. 
We find that $R_I^{-1}$ behaves very differently in the two regimes discussed in Section \ref{subsec:qualres}: for lower $R_0^{-1}$, $R_I^{-1} \simeq R_0^{-3/2}$, whereas $R_I^{-1}$ asymptotes to a constant for larger $R_0^{-1}$. The value of that constant is systematically close to $R_c^{-1}$, which would imply that the interfaces in that regime are just marginally unstable to ODDC. Why this is the case remains to be determined. However, \citet{Moll2016} showed that the turbulent transport through an ODDC-unstable region with an inverse density ratio close to $R_c^{-1}$ is essentially negligible (especially in the limit where $\Pr \sim \tau \rightarrow 0$), which explains why $\gamma^{-1} \simeq \gamma^{-1}_{\rm diff}$ even though the interfaces are not strictly quiescent. 

Going back to the results presented in Figure \ref{fig:gamma}, we are now able to explain the observed value of $\gamma^{-1}$ for high $R_0^{-1}$ runs. Indeed, since  $\gamma^{-1} \simeq \gamma^{-1}_{\rm diff} = \tau R_I^{-1}$, and since we find empirically that $R_I^{-1} \simeq R_c^{-1}$, we have 
\begin{equation}
\gamma^{-1} \simeq \tau R_c^{-1} = \tau \frac{\Pr + 1}{\Pr + \tau}  \equiv \gamma^{-1}_{c,{\rm diff}} 
\end{equation}
in the strongly stratified parameter regime where $R_0^{-1} \rightarrow R_c^{-1}$. The values of $\gamma^{-1}_{c,{\rm diff}}$ are shown as horizontal solid lines in each panel of Figure \ref{fig:gamma}, {\color{red} and predict the data relatively well}. In fact, we see that $\gamma^{-1}_{c,{\rm diff}}$ serves as an absolute minimum for the flux ratio, namely $\gamma^{-1}  \ge \gamma^{-1}_{c,{\rm diff}}$ in all simulations. 

Our findings have important consequences for core erosion models. Using the recent work of \citet{Soubiranal2017}, we can infer $\nu$ and $\kappa_C$ from their Figures 7 and 8 assuming a mean temperature $T_m \sim 30,000$K near the core-envelope interface, and a mean pressure $p_m \sim 30$GPa, to be roughly $\nu \simeq \kappa_C \simeq 2 \times 10^{-3}$cm$^2$/s (variations of $T_m$ and $p_m$ over the course the evolution of Jupiter do not affect our very rough estimates for $\nu$ and $\kappa_C$ significantly). The thermal diffusivity on the other hand can be obtained from \citet{Frenchal2012}, as $\kappa_T \simeq 2 \times 10^{-1}$cm$^2$/s. As a result, we have $\Pr \sim \tau \sim 0.01$. This would imply $\gamma^{-1}  \ge 0.5$, which is at least 5 times larger than estimated by \citet{Stevenson1982} and \citet{Guillot2004}.

\section{Models of Jupiter with the new flux ratio}
\label{sec:Jupitermodels}

\begin{figure}[h]
\centerline{\includegraphics[width=10cm]{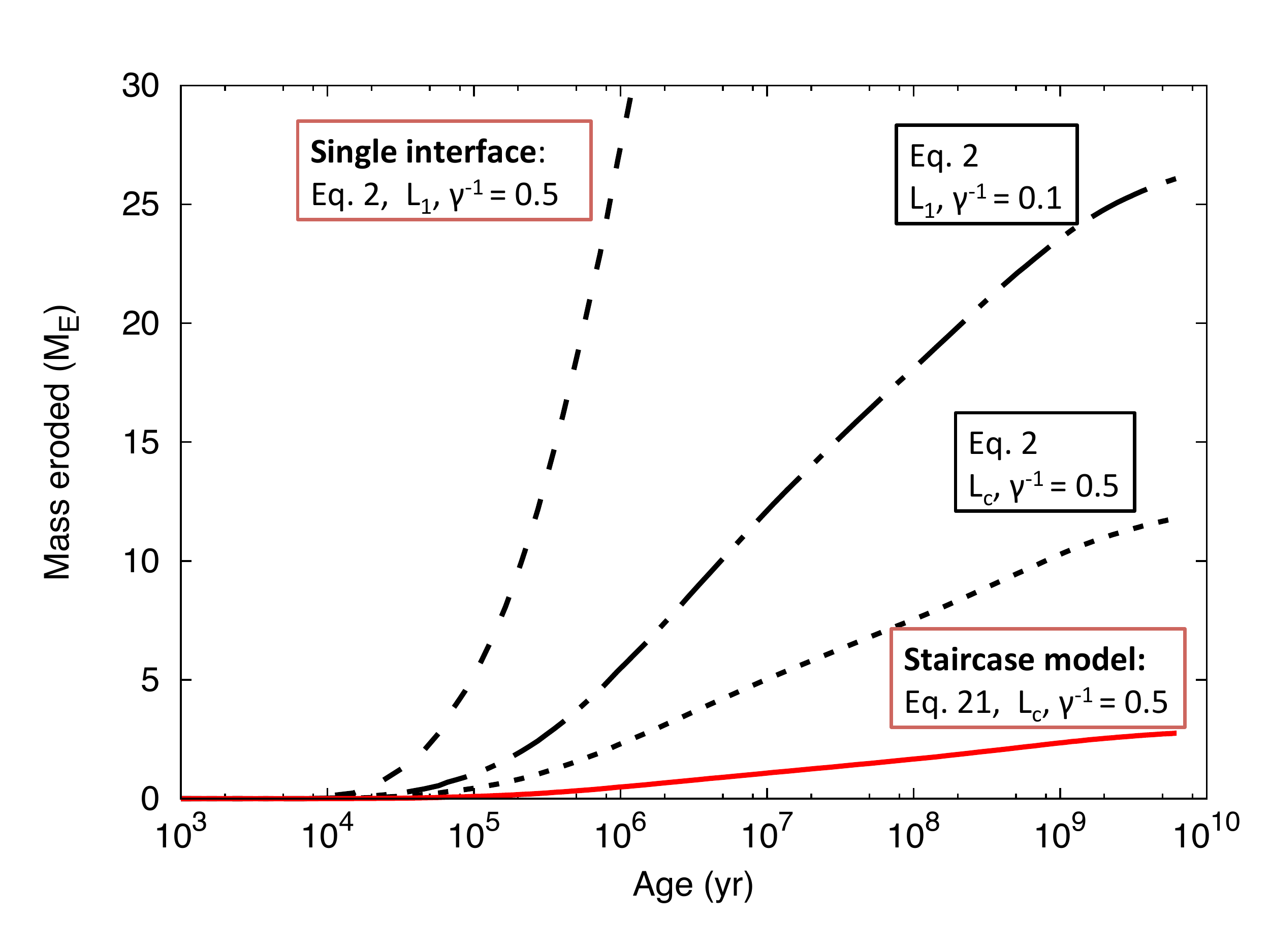}}
\caption{Core mass redistributed into the envelope as a function of time, using different values of $\gamma^{-1}$, of the relevant luminosity ($L_1$ one pressure scaleheight above the core or $L_c$ at the edge of the core) and different model equations: equation (\ref{eq:guillot}) derived by \citet{Guillot2004} and equation (\ref{eq:garaud}) proposed here. }
\label{fig:Jupiter}
\end{figure}

Figure \ref{fig:Jupiter} shows an estimate of the mass eroded away from Jupiter's core using the model of \citet{Guillot2004} (see equation \ref{eq:guillot}), with $\gamma^{-1} = 0.1$ and $\gamma^{-1} = 0.5$ respectively. The planet evolution code used to obtain these results is based on that of \citet{Thorngren2016}, with the addition of tabulated model atmospheres appropriate for Jupiter \citep{Fortney2011} and the water equation of state from \citet{French2009}. The evolution is carried out for a hot-start 1$M_J$ and 2$R_J$ model (where $M_J$ and $R_J$ are the observed mass and radius of Jupiter at the present day) having an isentropic hydrogen-helium envelope (with a helium mass fraction $Y=0.27$) surrounding an isothermal 30$M_E$ heavy-element core. Using $\gamma^{-1} = 0.1$  recovers the results of \citet{Guillot2004} with reasonable agreement (see their Figure 3.11), while using our new (lower-limit) estimate of $\gamma^{-1} = 0.5$ in the same model predicts that all of Jupiter's core should be eroded within the scope of 1Myr. 

Note, however, that equation (\ref{eq:guillot}) is derived from an order-of-magnitude energetic estimate (rather than being exact), assuming that a known fraction of the kinetic energy of the largest-scale eddies is used to lift heavy material from the core into the envelope. This is implicit for instance in the use of $L_1$, and of the term $GM/R$, which are both associated with convective eddies spanning a significant fraction of the entire convective envelope.  As such equation (\ref{eq:guillot}) should only be applied in the case where core erosion takes place through a single double-diffusive interface between the core and the envelope, as depicted in case 3 of Figure \ref{fig:cases}. On the other hand, if one assumes that erosion occurs through a multilayered double-diffusive staircase (cases 2 and 4 of Figure \ref{fig:cases}), then the transport processes are much more local and $\gamma^{-1}$ should relate the compositional flux across individual steps to the associated thermal flux, which can be estimated using the luminosity at the core-envelope interface $L_{\rm c}$. Starting with the definition of $\gamma^{-1}$ expressed as the ratio of non-dimensional buoyancy fluxes (see equation \ref{eq:gammadef}) and multiplying numerator and denominator by $\kappa_T |T_{0z}-T^{\rm ad}_z|$, we first obtain a ratio of dimensional fluxes, namely
\begin{equation}
\gamma^{-1} =  \frac{\beta (- \kappa_C  C_{0z} + \langle wC \rangle)  }{\alpha \left[ -\kappa_T (T_{0z}-T^{\rm ad}_z) + \langle wT\rangle \right]} \equiv \frac{\beta F_C}{\alpha (F_T + \kappa_T T^{\rm ad}_z)}, 
\end{equation}
where $F_T = - \kappa_T  T_{0z} + \langle wT \rangle$ is the local dimensional temperature flux, and $F_C = - \kappa_C  C_{0z} + \langle wC \rangle$ is the local {\color{red} dimensional} compositional flux.  Note that in most applications of interest, $|\kappa_T T^{\rm ad}_z | \ll F_T$, so we drop the diffusive term in $T_z^{\rm ad}$ in what follows. Then, moving back to a spherical geometry, and 
noting that $L_{\rm c} = 4\pi r_{\rm c}^2 \rho_{\rm c} c_{\rm p} F_T$ (where $r_{\rm c}$ is the core radius and $\rho_c$ is the local density at $r = r_c$), and $\frac{dm_{\rm core}}{dt} = - 4\pi r_{\rm c}^2 \rho_{\rm c} \beta F_C$, then 
\begin{equation}
\gamma^{-1} \simeq \frac{\beta F_C}{\alpha F_T } = \frac{4 \pi r_{\rm c}^2 \rho_c c_{\rm p} \beta F_C}{\alpha L_{\rm c}} = \frac{c_p}{\alpha L_{\rm c}} \left| \frac{dm_{\rm core}}{dt}  \right|  .
\end{equation}
This finally implies
\begin{equation}
\frac{dm_{\rm core}}{dt} \simeq - \gamma^{-1} \frac{\alpha L_{\rm c} }{c_{\rm p}}  .
\label{eq:garaud}
\end{equation} 
It is easy to verify that this equation is dimensionally correct. Figure \ref{fig:Jupiter} shows estimates of the core erosion rate using this formula with $\gamma^{-1} = 0.5$. Note that since the model is linear in $\gamma^{-1}$, the eroded core mass at a given time for other values of $\gamma^{-1}$ can easily be deduced from Figure \ref{fig:Jupiter}. Even with a large value of $\gamma^{-1}$, we see that the core erosion rate remains small, suggesting that most of Jupiter's core would be preserved when surrounded by a double-diffusive staircase. This conclusion is in part due to the use of the local luminosity $L_c$ (which is much smaller than $L_1$) as means to estimate the local buoyancy flux due to temperature, but not entirely. Indeed, using $\gamma^{-1} = 0.5$ and $L_c$ in the \citet{Guillot2004} model {\color{red} (equation \ref{eq:guillot})} still yields a substantially larger erosion rate than with the model proposed in equation (\ref{eq:garaud}).

\section{Conclusion}
\label{sec:conclusion}

In this paper we set out to study a possible scenario for the evolution of Jupiter, in which the planet is initially formed by core accretion, leaving it with a differentiated core of substantial mass. This core is later gradually eroded by convective motions, as suggested by \citet{Stevenson1982}. Except for the case of a solid core surrounded by a fully convective envelope (case 1 in Figure \ref{fig:cases}) the core-envelope region must be double-diffusive in nature, and can either take the form of a single interface (case 3), or a multilayered staircase (cases 2 and 4). We quantified, using DNSs, the ratio $\gamma^{-1}$ of the buoyancy flux associated with heavy elements transport to that associated with thermal transport across one or more double-diffusive interfaces, in order to improve estimates of the core erosion rate. We found that $\gamma^{-1}$ is substantially larger than previously thought, and appears to have a {\it lower limit} of $\tau (\Pr + 1) / (\Pr+\tau)$ (which is close to 0.5 in the interiors of giant planets), instead of the commonly-used estimate $\tau^{1/2}$ (which is closer to 0.1). This five-fold increase in $\gamma^{-1}$ implies a five times larger amount of material eroded from the core at any given age, given a particular core erosion model. Using the model of \citet{Guillot2004} (see equation \ref{eq:guillot}) with $\gamma^{-1} = 0.5$ predicts that the core would be completely eroded within 1Myr. This model is only appropriate if there is a single fluid interface (case 3), but the conclusion is consistent with that of \citet{Soubiranal2017} in the case of a single solid interface (case 1). We also proposed an alternative model in the presence of a developed, statistically stationary staircase (which is relevant for cases 2 and 4), and found that the opposite then happens: the erosion rate is very small throughout the planet's evolution and the core essentially remains close to its primordial size.

We therefore see that while the flux ratio $\gamma^{-1}$ across a double-diffusive interface is now fairly well constrained thanks to our DNSs, the main uncertainty in the core erosion rate is the choice of the core erosion model itself, which depends on the scenario one believes to be relevant (cases 1, 2, 3 or 4). Furthermore, model uncertainties remain even in the context of a given scenario. Indeed, the single interface core erosion model of \citet{Guillot2004} is based on simple energetic arguments, and has not yet been verified in the light of laboratory experiments or numerical simulations. While plausible on dimensional grounds, it could be off by factors of order unity, and such factors could change the conclusions reached. The staircase core erosion model proposed in equation (\ref{eq:garaud}) of this paper is also subject to significant uncertainties. Indeed, it implicitly assumes that the staircase is in global equilibrium, with equidistant steps that do not undergo mergers or splitting events. Allowing for such events (which are overall fairly likely) could invalidate our model setup, and would likely increase the transport through the staircase. This could, in turn, also change the conclusions reached. 

These uncertainties notwithstanding, we still believe that our work arrives at two important and robust conclusions. Firstly, that $\gamma^{-1}$ is much larger in the low Prandtl number astrophysical regime than in the high Prandtl number geophysical regime, and must be at least equal to 1/2. And secondly, that the core erosion rate seems to depend at least as much on what the nature of the core-envelope interface is (single interface vs. staircase) than on $\gamma^{-1}$ itself, with strong suggestions that the core would be entirely eroded by the present day if a single interface is present, while the latter could instead be preserved by a long-lived double-diffusive staircase. 

Over the lifetime of the Juno mission, data will continually accrue to help refine Jupiter's gravity field measurements. This will be used to construct revised interior structure models that will elucidate the case for an eroded core in Jupiter today. Similar gravity field data is now also being obtained for Saturn via the Cassini Mission's ``Grand Finale Orbits," which will be an important opportunity for comparative planetary science. A better understanding of double-diffusive transport across an interface or a staircase will also inform planetary evolution models, which in turn may help us constrain their formation history in the light of present-day observations. Indeed, assuming a single interface, the dredge up of core material would have a profound effect on the thermal evolution of Jupiter, as bringing up this dense material from the core would come at the expense of gravitational potential energy, which would lead to faster cooling than homogeneous models \citep{Fortney2011}. In the presence of a staircase, however, transport could be significantly reduced and cooling would slow down \citep{Stevenson1985,LeconteChabrier2012}. Finally, another essential open question regarding the formation of these planets is whether the measured enrichment in the heavier elements in the envelopes of both planets is due to the erosion of core material into the overlying envelope, or due to the accretion and ablation of planetesimals at young ages \citep[e.g.][]{Atreya2016}. How this enrichment comes about has taken on new significance now that exoplanetary metal enrichment has become an important area of study \citep[e.g.][]{Thorngren2016,Mordasini2016}.

\acknowledgments

The authors thank C. Caulfield, M. Davies-Wykes, D. Stevenson and T. Radko for interesting discussions. R. M. and P. G. were funded from NSF AST 1412951. This research was performed while R. M. held a grant from the National Research Council Research Associateship Program. C. M. acknowledges support from NASA Headquarters under the NASA Earth and Space Science Fellowship Program (grant NNX15AQ62H). The simulations were performed on the Hyades supercomputer, purchased using an NSF MRI grant.
Figure \ref{fig:snaps} was rendered using VisIt, a product of the Lawrence Livermore National Laboratory.


\end{document}